\documentclass[twocolumn,prl,sort&compress,floatfix]{revtex4-1}

\usepackage{amssymb}
\usepackage{amsmath}
\usepackage{mathrsfs}
\usepackage{enumitem}
\usepackage{cleveref}
\usepackage{bm}
\usepackage{tabularx}
\usepackage[dvipsnames]{xcolor}
\usepackage[sort&compress]{natbib}
\usepackage{graphicx}
\usepackage{color}
\usepackage{xr}
\usepackage{soul}

\usepackage{nonfloat}

\newif\iflong
\longtrue 

\begin{document}

\title{Modelling the 3D spatiotemporal organisation of chromatin replication}

\author{G. Forte$^1$, S. Buonomo$^2$, P. R. Cook$^3$, N. Gilbert$^4 $, D. Marenduzzo$^1$, E. Orlandini$^5$}
\affiliation{$^1$ SUPA, School of Physics and Astronomy, University of Edinburgh, Peter Guthrie Tait Road, Edinburgh, EH9 3FD, UK\\  
            $^2$ Institute of Cell Biology, School of Biological Sciences, University of Edinburgh, Edinburgh EH9 3FF\\  
            $^3$ Sir William Dunn School of Pathology, University of Oxford University, South Parks Road, Oxford OX1 3RE, UK \\
            $^4$ MRC Human Genetics Unit, MRC Institute of Genetics and Cancer, University of Edinburgh, Western General Hospital, Edinburgh EH4 2XU, UK \\
            $^5$ Dipartimento di Fisica e Astronomia and Sezione INFN, Universit\'a di Padova, Via Marzolo 8, Padova 35131, Italy.}
                



\begin{abstract}
We propose a polymer model for the dynamics of chromatin replication in 3D. 
Our simulations indicate that both immobile and tracking replisomes may self-assemble during the process, reconciling previous apparently discordant experimental evidence in favour of either scenario. Which of the two morphologies appears in our model depends on the balance between non-specific and origin-targeting interactions between chromatin and firing factors -- polymerases and other components of the replisome. Non-specific interactions are also necessary to yield clustering of factors and replication forks, creating structures akin to the replication foci observed in mammalian cells {\it in vivo}. 
We suggest that cluster formation provides an underappreciated but robust pathway to avoid stalled or faulty forks, which would otherwise diminish the efficiency of the replication process. Additionally, our simulations allow us to predict different modes of cluster growth during S-phase, which could be tested experimentally, and they show that 3D chromatin context is important to understand replication patterns in fission yeast. 
\end{abstract}
    
\keywords{chromatin replication, 3D polymer models, replication factories, chromatin damage, replication patterns}



\maketitle

\section{Introduction}

\begin{figure}[t]
\centering
\includegraphics[width=1\columnwidth]{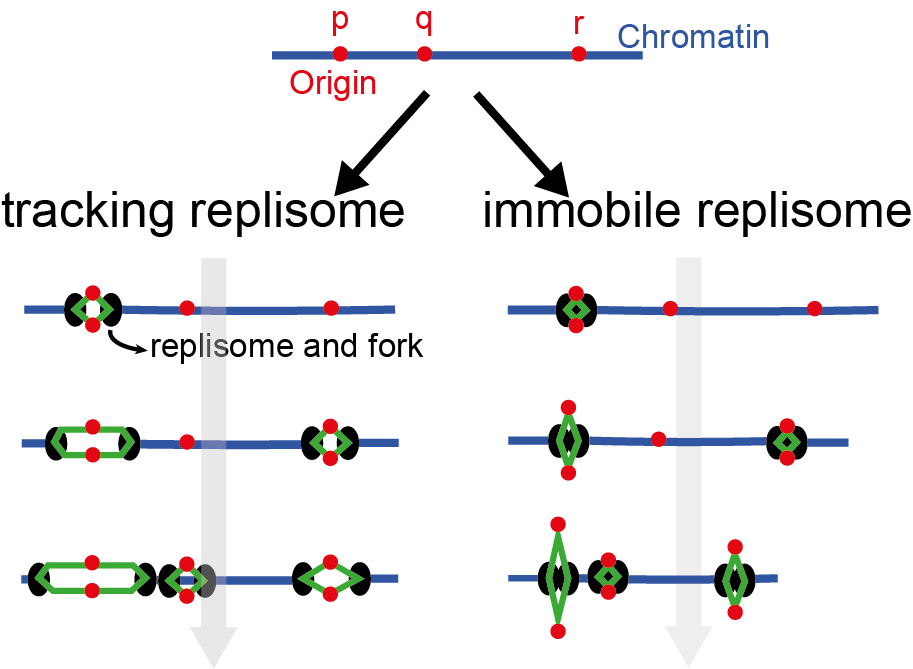}
\caption{\textbf{Replication by tracking or immobile replisomes.} In the tracking model (left), two replisomes bind to origin $p$, create a replication bubble, and move apart as they generate two new double helices (green segments); the process is repeated at $q$ and $r$. If immobile (right), two replisomes bind to origin $p$ and remain together as each pulls in template DNA and extrudes two new daughter helices; as before, the process repeats at $q$ and $r$. Note that each replisome in a pair is immobile relative to its partner, but pairs still move relative to other pairs. In both scenarios, the transparent grey arrow indicates time evolution.}\label{Fig:replisome_models} 
\end{figure}

Replication of eukaryotic DNA and chromatin is a crucial process in a cell's lifecycle. Whilst is normally depicted in 1 dimension (1D), it is inherently a three-dimensional (3D) process, in which spatio-temporal organisation is important~\cite{Alberts2017}. 
Two main models describe such organisation~\cite{Cook1999,Bates2008}. In the {\it tracking} model (Fig.~\ref{Fig:replisome_models}, left), two replisomes loaded on one origin move away from each other as they replicate. In the {\it immobile replisome} model (Fig.~\ref{Fig:replisome_models}, right), the two are in constant contact, as template DNA is pulled in from each side and two double-stranded loops are extruded. There is evidence for and against both models. Thus, on one hand, fork progression {\it in vitro} is unaffected by omitting from the reaction the Ctf4 molecules that hold the two helicases together in the pre-replication complex~\cite{Yeeles2015} and single-molecule imaging of extracts from {\it Xenopus} eggs reveals individual replisomes tracking independently along templates~\cite{Yardimci2010}. On the other hand, structures of replication complexes assembled {\it in vitro} are consistent with a central and immobile dimer that extrudes daughter duplexes~\cite{Yuan2019}, and more evidence for the immobile replisome model is reviewed in~\cite{Li2020}. Experiments on bacteria also suggest that both models may apply in a single cell, as replisomes are at times moving together, and at others tracking independently~\cite{Chen2023}. 

Besides being an integral part of the immobile replisome model, clusters (of polymerases and replisome elements) are also observed at a higher organizational level~\cite{Xiang2018}. Thus, many human replisome pairs have been found to form clusters, called replication \textit{foci} or \textit{factories}. Such clusters are usually small at the beginning of the S phase, before enlarging and changing nuclear position~\cite{Vouzas2021, Chagin2016}. Notably, mechanisms leading to the change in foci size remain unclear. For instance, early experiments suggested that mammalian foci are fixed in 3D space and the clustering observed during S-phase is due to continuous disassembly and reassembly of whole replisomes~\cite{Anachkova2005}, whilst a more recent work shows mobile yeast foci continuously fusing and segregating~\cite{Meister2007}. 

The replication fork -- the DNA sites where a helicase and polymerases are working together to replicate the genomic material -- moves at a speed that depends on the  organism, with an average of $1.6-3 \,$ kb/min in yeast~\cite{Gispan2017}. Forks are normally thought to be asymmetric, such that the leading strand is synthesised almost continuously, whereas the lagging strand is replicated by stitching together short Okazaki fragments, and the polymerase on the lagging strand disengages often. Additionally, obstacles such as RNA polymerases or DNA damage can slow or stall forks -- on either the leading and lagging strand -- and this can induce replication stress and the development of common fragile sites~\cite{Berti2016, Franchitto2013}. 
Recent single-molecule photobleaching experiments show that most components in bacterial and yeast replisomes (even leading-strand polymerases) exchange rapidly and continuously with the soluble pool~\cite{Beattie2017}, and this could facilitate the progress of a fork halted by a blockage. 
Consequently, one may imagine that the system has to strike a compromise between the tight binding required to keep replisomes on the DNA, whilst allowing a sufficiently dynamic exchange to avoid stalling. 



Whilst many models describing replication dynamics and origin firing have been developed (see, e.g.,~\cite{Shaw2010,Retkute2012,Gindin2014,Das2015,Lob2016,Arbona2018,Jun2006}), most are inherently 1D and very few include the critical role played by 3D effects and polymer physics. Additionally, we are aware of no 3D polymer model which investigates the mechanism controlling the dynamics of polymerase or fork clusters in chromatin. 
Modelling spatial patterns in eukaryotic replication is complicated because newly replicated polymer with its steric hindrance has to be dynamically created; additionally, one has to account for the binding-unbinding dynamics of firing factors - identified here as polymerases and other replisome elements --, which is not well characterised experimentally. Nevertheless, this type of modelling has the potential to generate new hypotheses to be tested experimentally, which go beyond the prediction of firing efficiency and replication profiles along the 1D genome. \\
To address this gap, here we develop a 3D polymer model to study the dynamics of chromatin replication and, in particular, of clusters of firing factors and forks (i.e., replication factories/foci).
This model is minimal in the sense that as few parameters as possible have been retained to describe this complex system; yet the model can both reproduce some key properties of eukaryotic replication and also give some predictions about key ingredients for the spatial organisation of replication. First, we show that in simulations with multiple origins both the immobile and the tracking replisome models can be observed in 3D, and the balance between the two depends on the ratio between non-specific interactions between firing factors and chromatin, and specific interactions between factors and origins. The presence of non-specific interactions is key in the simulations to observe the immobile replisome scenario, and in general for the formation of clusters of factors and origins. 
These results lead to the prediction that the emerging clusters are likely to have a functional role, as they fuel the restarting of replication when polymerases are temporarily lost, for instance after replication of an Okazaki fragment, thereby diminishing the likelihood of stalled forks, which would hinder the efficiency of the replication process. Second, simulations suggest different scenarios for the spatiotemporal evolution of the replication foci/factories. Thus, clusters in our simulations may grow by collisions or by a new and unexpected formation of long-range chromatin loops.  Finally, simulations yield replication patterns that mimic those observed experimentally in wild-type fission yeast and a Rif1 mutant that up-regulates firing frequencies of some origins whilst down-regulating others~\cite{Hayano2012}.

\section*{Results}\label{sec2}
\subsection*{A 3D polymer model for chromatin replication}

\begin{figure}[!]
    \centering
    \includegraphics[width=1\columnwidth]{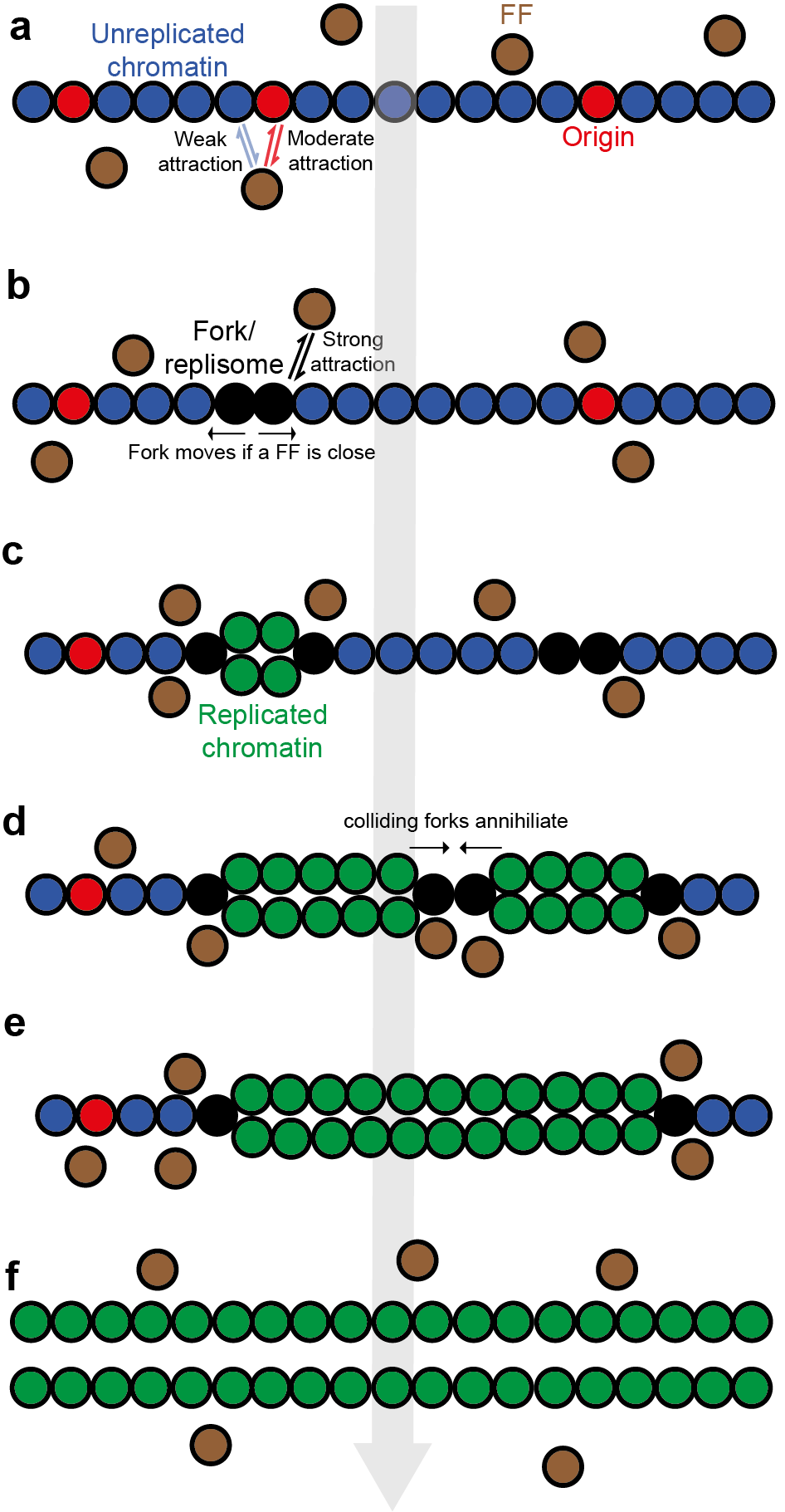}
    \caption{\textbf{3D polymer model for chromatin replication.} Chromatin is represented as a polymer composed of several beads, each one corresponding to  $1 \, kbp$. \textbf{a} The initial chromatin filament contains unreplicated chromatin sites (blue beads) and replication origins (red beads). Firing factors (FFs, brown spheres)  bind weakly (non-specifically) to unreplicated chromatin sites and strongly to replication origins. \textbf{b} When an FF binds to an origin, the origin \textit{fires} with probability $P_{fire}$; this involves converting the origin plus one of its randomly-chosen neighbours into two forks or replisomes (black beads) which experience a very strong attraction with FFs.  \textbf{c} The two forks independently move in opposite directions whenever a FF is close by. This bi-directional replication results in the formation of two chromatin filaments (green beads) which experience a steric interaction with FFs.  \textbf{d-e} The collision of two forks results in their annihilation and the replicated filaments generated by each fork join together. \textbf{f} Replication ends when the whole string is 'replicated' into two green strings.}\label{Fig:model}
\end{figure}

In this study, we develop a coarse-grained model to study replication at the chromatin level, without specifiying the molecular details about the replication machinery. Critically,  our new model is based on as few assumptions as possible, such that it could be applied throughout eukaryotes despite considerable variations in origin size, spacing, and firing frequency~\cite{Hu2023, Wang2021, Zhao2020, Rhind2022}. 
Chromatin is depicted as a semi-flexible polymer composed of a sequence of beads connected by springs (Figure~\ref{Fig:model}). Each bead is assumed to have a diameter  $\sigma = 15 \, nm$ and contains $1$ kbp of DNA~\cite{Chiang2022}. An additional potential among triplets of consecutive beads provides a persistence length $l_p \sim 60 \, nm$, roughly that of chromatin~\cite{Langowski2006,Langowski2007} (see Material and Methods for detail). 
A chromatin filament initially contains two types of sites: unreplicated chromatin sites (blue beads in Figure~\ref{Fig:model}) and replication origins (red beads in Fig.~\ref{Fig:model}). All proteins required for replication are represented by brown spheres which we call 'firing factors' (FFs), and include all components necessary to complete a whole replication cycle (i.e., activating kinases, MCM proteins, helicases, polymerases replicating leading and lagging strands, topoisomerases, and termination proteins). Importantly, the concentration of FF in our simulations is limiting, meaning that only a few origins in a multi-origin chromatin filament can be active at any time, as observed \textit{in vivo}~\cite{Fragkos2015}. \\
FFs diffuse throughout space whilst being excluded from the volume occupied by all other beads (Fig.~\ref{Fig:model}). Their multivalent binding to different types of chromatin beads is modelled by a truncated and shifted Lennard-Jones potential 
\begin{equation}
V_{LJ/cut} (r_{i,j}) = \left[ V_{LJ}(r_{i,j}) - V_{LJ}(r_c) \right] \Theta(r_c-r_{i,j})\label{Eq:LJ_potential},
\end{equation}
where $r_{i,j}$ is the distance between the $i-th$ and $j-th$ beads, $r_c = 1.8 \, \sigma$ is a cut-off distance and
\begin{equation}
V_{LJ} (r) =  4 \varepsilon \left[ \left(\frac{\sigma}{r_{i,j}} \right)^{12} - \left(\frac{\sigma}{r_{i,j}} \right)^{6} \right], \label{Eq:LJ}
\end{equation}
where $\varepsilon$ is the interaction energy. FFs are assumed to be weakly (non-specifically) attracted to (blue) unreplicated chromatin beads ($\varepsilon = \varepsilon_{ns} = 4 k_BT$ in Eq.~\ref{Eq:LJ}) and moderately to (red) origins ($\varepsilon= \varepsilon_{origin}= 6 \, k_BT$ in Eq.~\ref{Eq:LJ}). The importance of a weak interaction between FFs and unreplicated chromatin beads is explained below and represents a key constituent of our model. 
Replication is modelled as follows. Once a FF binds to an origin (Figure~\ref{Fig:model}a), the latter fires with probability $P_{fire}$ to create a pair of replisomes or forks (two black beads derived from the red bead plus a randomly-chosen blue neighbour; Fig.~\ref{Fig:model}b). Each fork binds FFs strongly ($\varepsilon = \varepsilon_{fork} = 10 \, k_BT$ in Eq.~\ref{Eq:LJ}) and moves independently and in the opposite direction to its partner, provided that a FF is within $r_{c}=1.8 \, \sigma$. 
The fork movement along the template chromatin strand results in the replication of the template strand itself and in the formation of a new strand (green beads in Figure~\ref{Fig:model}). Both strands are identical with respect to their biophysical properties and only interact through steric repulsion with FFs. During the simulation multiple origins can fire and, when two forks travelling in opposite directions collide (Fig.~\ref{Fig:model}d), they annihilate each other to leave appropriately connected replicated chains (Fig.~\ref{Fig:model}(d-e)). Through successive origin firing events and fork movements, the original blue/red chain is replaced by two replicated green chains which become disconnected at the end of replication (Fig.~\ref{Fig:model}f). The final two replicated filaments cannot be re-replicated (in accord with what is seen{~\it in vivo}~\cite{Cook2001}) as they contain no 'licensed' origins, and as FFs have no affinity for green replicated beads. \\
The entire system is subject to Brownian dynamics which is integrated through the software LAMMPS~\cite{Plimpton1995}.

\subsection*{Replicating a chromatin fibre with one origin}

\begin{figure}[t]
\centering
\includegraphics[width=0.9\columnwidth]{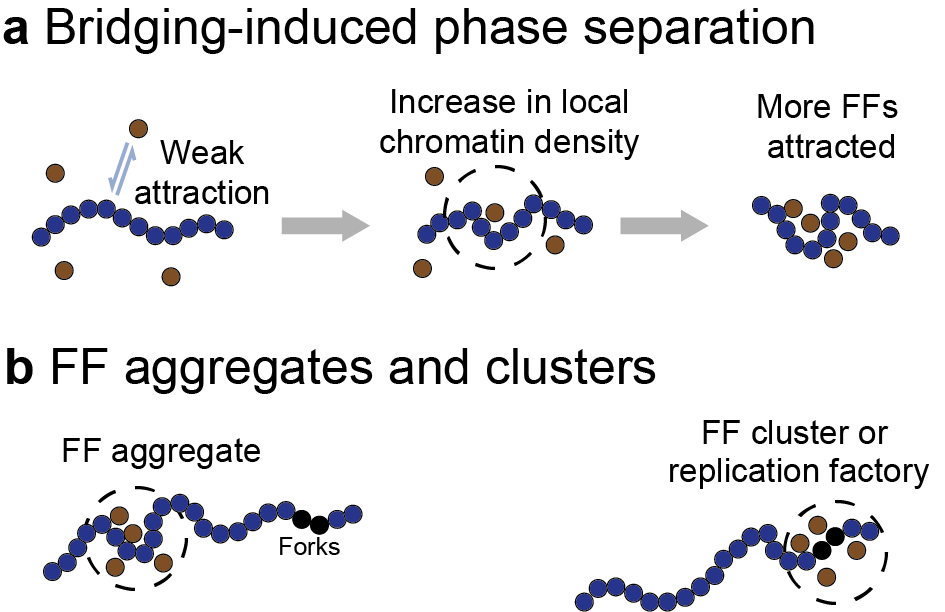}
\caption{\textbf{BIPS mechanism and aggregate/cluster definition.} \textbf{a} Left: the bridging-induced phase separation (BIPS) is driven by the presence of a weak attraction between the chromatin filament (blue chain) and firing factors, FFs (brown spheres). Center: An FF binds the chromatin filament and, because of its multivalent nature, it can bind multiple beads composing the chain resulting in a local increase in chromatin density. Right: the larger chromatin density attracts even more FFs which form agglomerates, even without being attracted to each other. \textbf{b} Left: we define FF aggregates as those FF agglomerates that are not localised close to replication forks. Right: FF clusters (or equivalently, replication factories) are defined as FF agglomerates in the proximity of forks. Both aggregates and clusters form because of the BIPS mechanism.} 
\label{Fig:sketch}
\end{figure} 

\begin{figure}[htbp]
\centering
\includegraphics[width=0.9\columnwidth]{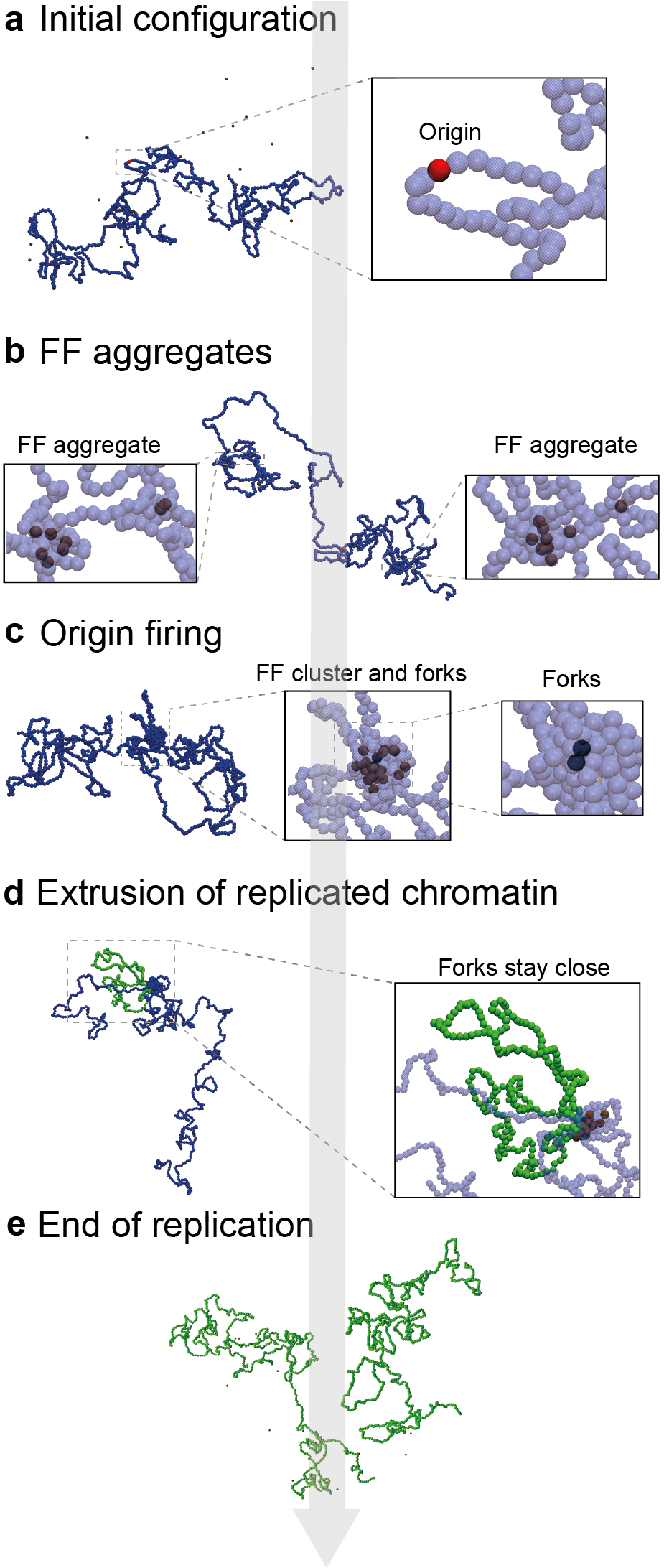}
\caption{\textbf{Frames from a simulation about replication of a chromatin filament with a single origin.} \textbf{a}  Initial configuration containing the relaxed chromatin filament and diffusing FFs. The chromatin polymer is composed of $1000$ beads ($1 \, Mbp$) and the origin is placed in the centre. Blue beads in the inset are represented as transparent to highlight the origin (the same representation will be used in the following panels). \textbf{b} After switching on the attraction between FFs and chromatin, FF aggregates form. \textbf{c} When an FF aggregate gets close to the origin, the latter fires with probability $P_{fire}=0.01$, resulting in the formation of two forks and the FF aggregate is now defined as FF cluster or replication factory/foci. \textbf{d} Replicated chromatin is extruded by the two forks that stay close in $3D$ throughout the whole replication process. \textbf{e} The simulation ends when the whole initial chromatin filament has been replicated, leaving two independent fibres. } 
\label{Fig:1_origin}
\end{figure} 

We start by analysing the replication of a chromatin filament formed by $1000$ beads (equivalent to $1 \, Mbp$), with a single origin in the middle, and surrounded by $20$ firing factors. The origin fires with probability $P_{fire}=0.01$. Unless specified otherwise, this firing frequency will be used throughout as it is close to the median of $0.037$ obtained for human initiation zones~\cite{Wang2021}, and because it gives sufficiently slow dynamics to be biophysically realistic, whilst remaining tractable in terms of computational time.\\
Before presenting the simulation results of our model, it is necessary to make the following clarifications. As we will show, in the simulations FFs are observed to form agglomerates. Notably, this phenomenon occurs without introducing any specified attractive interaction or cooperation between FFs but is only due to a positive feedback mechanism known as the bridging-induced phase separation (BIPS) effect that depends on the multivalent bindings of FFs on the chromatin and which works as follows~\cite{Brackley2013,Brackley2020}:  the initial non-specific binding of a FF to a chromatin segment results in an increase in the local chromatin density that, in turn, attracts more FFs giving rise to FF agglomerates (see Fig.~\ref{Fig:sketch}a).\\
The FF agglomerates can be partitioned into two subsets:   those that are spatially close to more than one fork (\textit{clusters}) and all the others that are instead far away from the replication forks (\textit{aggregates}), see Fig.~\ref{Fig:sketch}b). In this context, a cluster is analogous to a replication factory or replication foci. Note that, within the immobile replisome scenario,  such clusters can arise also if a single origin is present as two forks colocalise.\\
We now go into the details of the replication of a chromatin chain with a single origin. A simulation (Movie S1) begins with a relaxed chain and diffusing (but non-binding) FFs (Fig.~\ref{Fig:1_origin}a). As soon as the attraction between the latter and the chromatin filament begins, a spontaneous formation of FF aggregates is observed changing the configuration of the chain before the start of replication (Fig.~\ref{Fig:1_origin}b).
Eventually, in the setup of Fig.~\ref{Fig:1_origin}, all FFs localise in the proximity of the replication origin (where the affinity is larger) which fires with probability $P_{fire}=0.01$. The firing event generates a pair of forks/replisomes surrounded by the FF aggregate which, in line with the previous definition (Fig.~\ref{Fig:sketch}b), can be identified as an FF cluster or replication factory (Fig.~\ref{Fig:1_origin}c). Strikingly, the resulting forks remain close in 3D space  (Fig.~\ref{Fig:1_origin}d) despite no forces between them being specified, until the end of the simulation where two replicated strands freely diffuse independently from each other (Fig.~\ref{Fig:1_origin}e).
This simple model therefore captures several elements of replication {\it in vivo} -- mainly, the extrusion of two daughter loops by a replisome pair, and the maintenance of contact between the two replisomes in a pair. 
Importantly, the key element to maintaining the two replisomes together is the presence of a weak non-specific attraction and the consequent formation of the FF cluster, as we will discuss in the next session.

\subsection*{Non-specific interactions are required to keep replisomes together}

\begin{figure}[htbp]
    \centering
    \includegraphics[width=1\columnwidth]{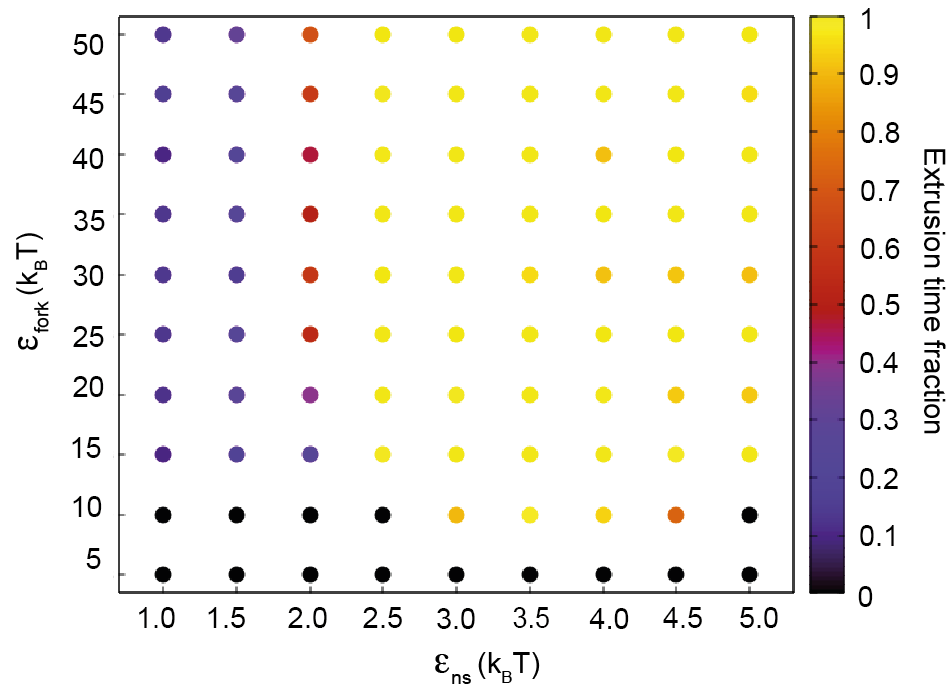}
    \caption{\textbf{Phase diagram for the extrusion time fraction in a chromatin filament composed by $1000$ beads with a single origin in the middle.} The phase diagram shows the extrusion time fraction, which is defined as the fraction of the replication time (colour scale on the right) during which extrusion of the replicated filaments is observed, depending on $\varepsilon_{ns}$ (x-axis) and $\varepsilon_{fork}$ (y-axis). The attraction between FFs and the origin, $\varepsilon_{origin}$,  is kept constant and equal to $8 \, k_BT$. The replication time fraction is quantified by monitoring the $3D$ distance between the two forks and assuming extrusion happens whenever this distance is smaller than $6 \, \sigma$, which is the maximum displacement between the two forks observed in Fig.~\ref{Fig:1_origin}. 
    For each  couple of parameters $(\varepsilon_{ns}, \varepsilon_{fork})$, the replication extrusion fraction is computed by averaging over $10$ independent simulations. Black dots correspond to cases when the replication process is so slow that the replication of the whole template chromatin filament takes too long computational time and the total replication time itself would be unrealistic. }
    \label{Fig:phase_diagram} 
     
\end{figure}

We now examine the role of the weak non-specific attraction, between FFs and unreplicated chromatin sites that are not origins (blue beads in Fig.~\ref{Fig:model}), in the formation of a replisome pair.  We begin by asking whether without non-specific attractions one can still observe the two forks close in $3D$ throughout the whole replication process. To address this question, we perform simulations where we keep the parameters of the model as previously ($\varepsilon_{origin}=6 \, k_BT$, $\varepsilon_{fork}=10 \, k_B T$), but now remove the non-specific attraction between (blue) unreplicated chromatin beads and FFs (this is achieved by using a Weeks-Chandler-Anderson potential, representing a purely steric interaction). The firing probability has here been increased to $P_{fire}=1$ to speed up simulations; the results are not qualitatively affected by this different value of $P_{fire}$. 
Importantly, without non-specific interactions, FF agglomerates or clusters are not observed, and -- once the origin fires -- the two replisomes separate over time (Movie S2). 

By increasing the value of $\varepsilon_{fork}$, it could be expected that eventually there would be  a scenario where a single FF binds the two forks so strongly to keep them together during the whole replication process, without requiring the presence of an FF cluster. However, this value corresponds to $\varepsilon_{fork,min} > 50 \, k_BT$, which is unrealistically large, as we now show. To compare with realistic affinities between forks and FFs, we can use statistical mechanics calculations to relate the Lennard-Jones potential in Eq.~\ref{Eq:LJ_potential} (and hence $\varepsilon_{fork}$) to the dissociation constant $K_D$, which is used more commonly in biochemical assays to determine the strength of a ligand-protein interaction, and which equals the concentration of ligands for which half proteins are bound~\cite{Berg2007}. Note that the smaller $K_D$ is, the stronger the attraction. 

The equation relating $\varepsilon_{fork}$ to $K_D$ is (see SI for details): 
\begin{equation}\label{statmech}
\frac{1}{K_D} = 4\pi \int_0^{r_c} x^2 \exp\left[-4\beta\epsilon_{fork} \left(\frac{1}{x^{12}} - \frac{1}{x^6}\right)\right]dx,
\end{equation}
where $r_c$ is the Lennard-Jones cutoff distance $r_c=1.8 \, \sigma$. The $K_D$ required to keep two forks together (corresponding to $\varepsilon_{fork,min} > 50 \, k_BT$)  is then sub-picomolar ($K_{D,min} << 1 \, pM$), far smaller than the smallest nanomolar dissociation constants found {\it in vivo}~\cite{Alberts2017}. We conclude that, in the absence of non-specific interactions between FFs and chromatin, replisome pairing requires unrealistically small dissociation constants, hence would not be observed, and replisomes would track independently. A phase diagram showing more quantitatively the fraction of the replication time during which the replisomes stay together is shown in Fig.~\ref{Fig:phase_diagram}. To obtain this phase diagram, the replication of a chromatin filament composed by $1000$ beads with a single origin in the middle (as the one in Fig.~\ref{Fig:1_origin}) was investigated by using different values of $\varepsilon_{ns}$ (x-axis) and $\varepsilon_{fork}$ (y-axis). It is possible to observe that at low $\varepsilon_{ns}$ extrusion is quite rare, independently from $\varepsilon_{fork}$. It instead becomes relevant when the non-specific attraction between FFs and unreplicated chromatin sites increases, as an FF cluster is formed and the two replisomes are kept together in $3D$.  For $\varepsilon_{fork}=10 \, k_BT$, there is a significant increment of the extrusion time fraction while $\varepsilon_{ns}$ increases from $1 \, k_BT$ to $4 \, k_BT$ as the FF cluster nucleates. For larger values of $\varepsilon_{ns}$, the extrusion time decreases because  $\varepsilon_{fork}$ becomes comparable to $\varepsilon_{ns}$, leading to a competition for FFs between forks and unreplicated chromatin.

The phase diagram in Fig.~\ref{Fig:phase_diagram}, together with the statistical mechanics calculation in Eq.~\ref{statmech}, demonstrate the requirement of non-specific interactions to observe replisome pairing concomitant with FF clustering. 

\subsection*{Replicating a chromatin fibre with multiple origins}

\begin{figure}[htbp]
    \centering
    \includegraphics[width=1\columnwidth]{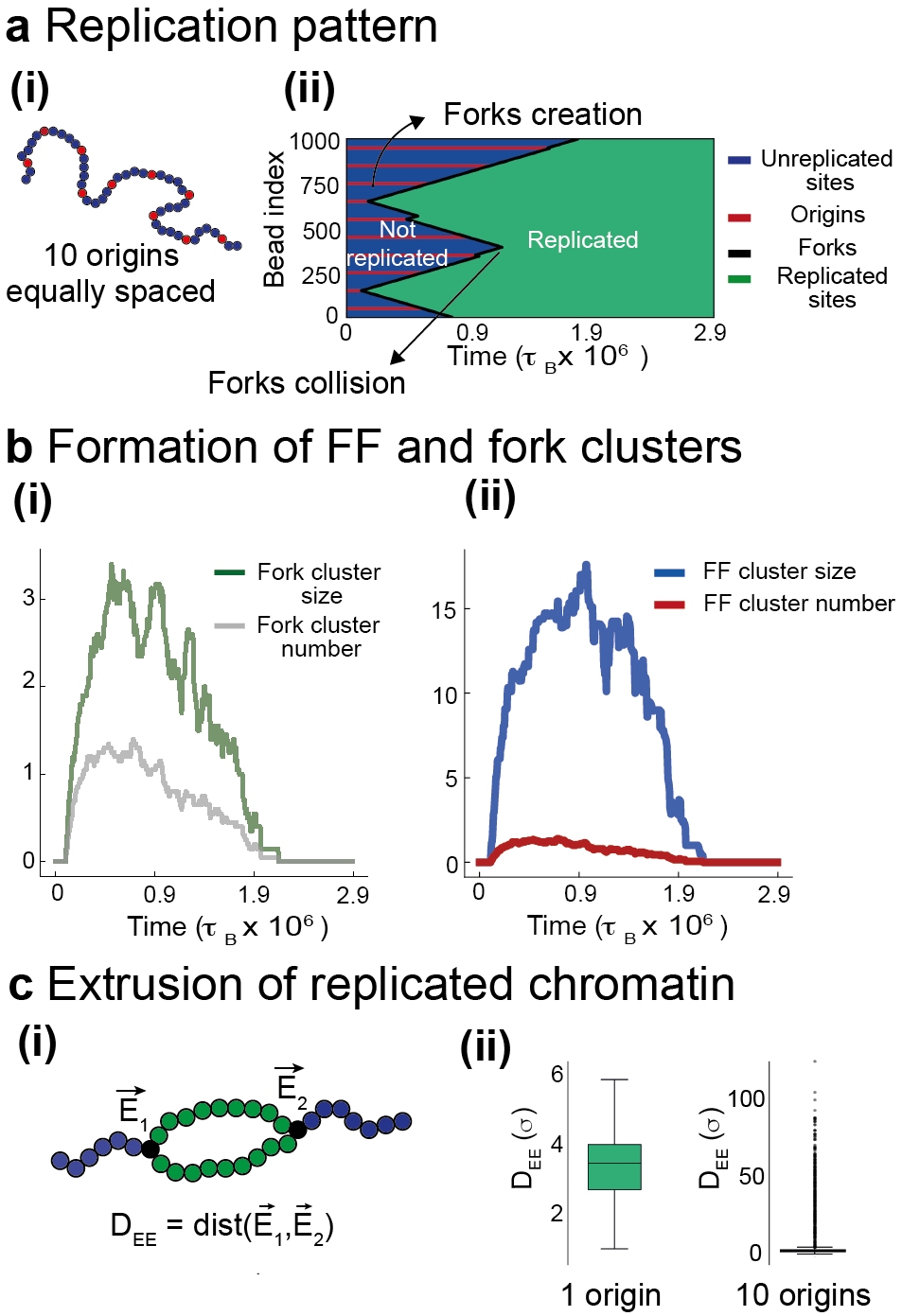}
    \caption{\textbf{Replication of a chromatin filament composed by $1000$ beads and containing $10$ equally spaced origins.} \textbf{a} Sketch of the chromatin polymer containing multiple origins (panel (i)) and an example of a kymograph (panel (ii)). In the kymograph, the y-axis provides the chromatin bead number (from $1$ to $1000$) and the x-axis provides the simulation time. Blue, red, black, and green pixels indicate the bead type (unreplicated sites, origins, forks, and replicated sites, respectively). Fork creations and collisions appear as green peaks and valleys. 
    \textbf{b} Changes over the replication time in the average size and number of clusters of forks (panel (i)), and in the size and number of clusters of FFs (panel (ii)). A cluster is composed by particles (forks or FFs) whose $3D$ distance is smaller than $4 \, \sigma$ for forks and $2 \, \sigma$ for FFs. The size and number of fork clusters do not appear to anti-correlate significantly, as a result of the small number of forks present in the system, and their constant creation and annihilation. FF clusters, instead, increase in size (blue curve) and decrease in number (red curve) as observed in microscopy experiments investigating the dynamics of PCNA complexes. Averages are from  $20$ independent simulations.
    \textbf{c} Panel (i): the realisation of the immobile or tracking replisome scenario is quantified by computing, at each time point, the $3D$ distance between the two extremities of each replicated chromatin fragment $D_{EE}$. Panel (ii): in the presence of a single origin (distribution on the left), $D_{EE}$ always assumes small values indicating the immobile replisome scenario as observed in Fig.~\ref{Fig:1_origin}. In the $10$ origins set-up (distribution on the right), the $D_{EE}$ median value is still small, but larger outliers are observed, indicating that tracking replisomes can also occur. }
    \label{Fig:multiple_origins} 
     
\end{figure}

As chromosomes in eukaryotes are often replicated from multiple origins, we next consider a chain composed of $1000$ beads (representing $1 \,$ Mbp) with $10$ equally-spaced origins (Fig.~\ref{Fig:multiple_origins}a(i)), each of them firing with probability $P_{fire}=0.01$. As in the previous paragraph, the system contains $20$ firing factors. In this set-up, multiple replisome pairs can be active at the same time (see Movie S3) and the temporal evolution of replication can conveniently be depicted using a kymograph, where, for each time point (x-axis), differently-coloured chromatin beads (whose index is reported in the y-axis) are represented by appropriately-coloured pixels (Fig.~\ref{Fig:multiple_origins}a(ii)). Origin firing and fork merging are marked by green peaks and valleys, respectively. In the plot, the formation of multiple fork pairs (like the ones originated by origins close to beads $250$ and $750$) can be observed, as well as the passive replication of origins that do not fire during the simulation (like the origin close to the bead $1000$). Such stochastic and infrequent firing is the norm in mammalian cells~\cite{Wang2021}

\subsection{Clusters of forks and firing factors form and grow spontaneously}

Previous microscopy experiments revealed the formation and growth of clusters (or replication factories/loci)  by tracking both elements of the replication machine~\cite{Leonhardt2000} and of replisomes or forks~\cite{Nakamura1986, Leonhardt2000}. In both cases, that the size of these clusters appears to increase in time, while their number decreases. 
To address such dynamics, the evolution of clusters of FFs and forks, which are the equivalent to replication factories/foci {\it in silico}, was analysed during the replication of a chromatin filament containing $10$ equally-spaced origins as in the previous section. It is important to remember that an FF cluster is defined as an FF agglomerate that is spatially close to more than one fork (Fig.~\ref{Fig:sketch}a), while a fork cluster is simply defined as an agglomerate of forks. To investigate the number and size of clusters, we make use of the algorithm provided in Ref.~\cite{Stoddard1978} and say that two particles (either two forks or two FFs) belong to the same cluster if their $3D$ distance is smaller than the threshold $r_{thre}= 4\, \sigma$ for forks and $r_{thre}=2 \, \sigma$ for FFs (small variations in $r_{thre}$ do not change the results). The cluster size is then defined as the number of particles composing the cluster.\\
Several considerations are provided by our simulations. 
First, FF clusters form in the presence of multiple origins as well as forks originating from different origins organised in clusters (see Movie S3). The underlying mechanism is again driven by BIPS. 
Second, Fig.~\ref{Fig:multiple_origins}b shows the number and size of forks and FF clusters averaged over $20$ independent simulations. We observe that the size of fork clusters oscillates significantly, due to the creation and annihilation of fork pairs during the replication process (Fig.~\ref{Fig:multiple_origins}b(ii)) and to the small number of forks generated by our $10$-origin system. At the same time, the number of fork clusters slowly decreases in the second half of the simulation, after an initial increase. A more striking connection with replication factories, seen by early microscopy studies~\cite{Cook1999}, is evident when we look at clusters formed by FFs, whose number is constant in time and is substantially larger than the average number of replication forks. Fig.~\ref{Fig:multiple_origins}b(ii) shows that, after a small increment, the number of clusters of firing factors (red curve) starts decreasing, while their size (blue curve) increases for a while before decreasing again toward the end of replication. The small increase in the number of FF clusters is due to the initial formation of multiple small clusters which quickly merge together, thus increasing in size. The coarsening of actively replicating clustering, in broad qualitative agreement with the dynamics of replication clusters seen by microscopy~\cite{Hozak1994}, is due to the combination of BIPS with the motor activity of FFs. 

\subsection*{Replisome pairs are marginally less stable in the multi-origin set-up}

We now ask whether, in a multi-origin set-up, replicated chromatin is often extruded by pairs of forks close in $3D$, as for the $1$-origin case (Fig.~\ref{Fig:1_origin}).
Extrusion can be quantified by considering each replicated chromatin filament in the system (at each time step) and computing the $3D$ distance between its two extremities at positions $\vec{E_1}$ and $\vec{E_2}$,  $D_{EE}= dist(\vec{E_1}, \vec{E_2})$ (see Fig.~\ref{Fig:multiple_origins}c(i)).  The distributions of $D_{EE}$ show substantial differences in the $1$-origin and multi-origin set-up (Fig.~\ref{Fig:multiple_origins}c(ii)). In the presence of a single origin, $D_{EE}$ remains small at all times ($1\, \sigma \lesssim D_{EE} \lesssim 6 \, \sigma$), indicating that the two forks always stay close to each other as we previously saw. In the multi-origin case, $D_{EE}$ can instead also assume very large values (even larger than $100 \, \sigma$), even if the median of the distribution remains low. The large values of $D_{EE}$ found for some configurations indicate that extrusion is not always observed and that a fork can even track independently along the chromatin filament (see Fig.S1). By analysing the fraction of chromatin segment configurations, the fraction of non-extruding forks, with the parameters of Fig.~\ref{Fig:multiple_origins}, is estimated to be very low, i.e. around $4\%$.

\subsection*{Mechanisms driving cluster growth}

\begin{figure}[t]
    \centering
    \includegraphics[width=1\columnwidth]{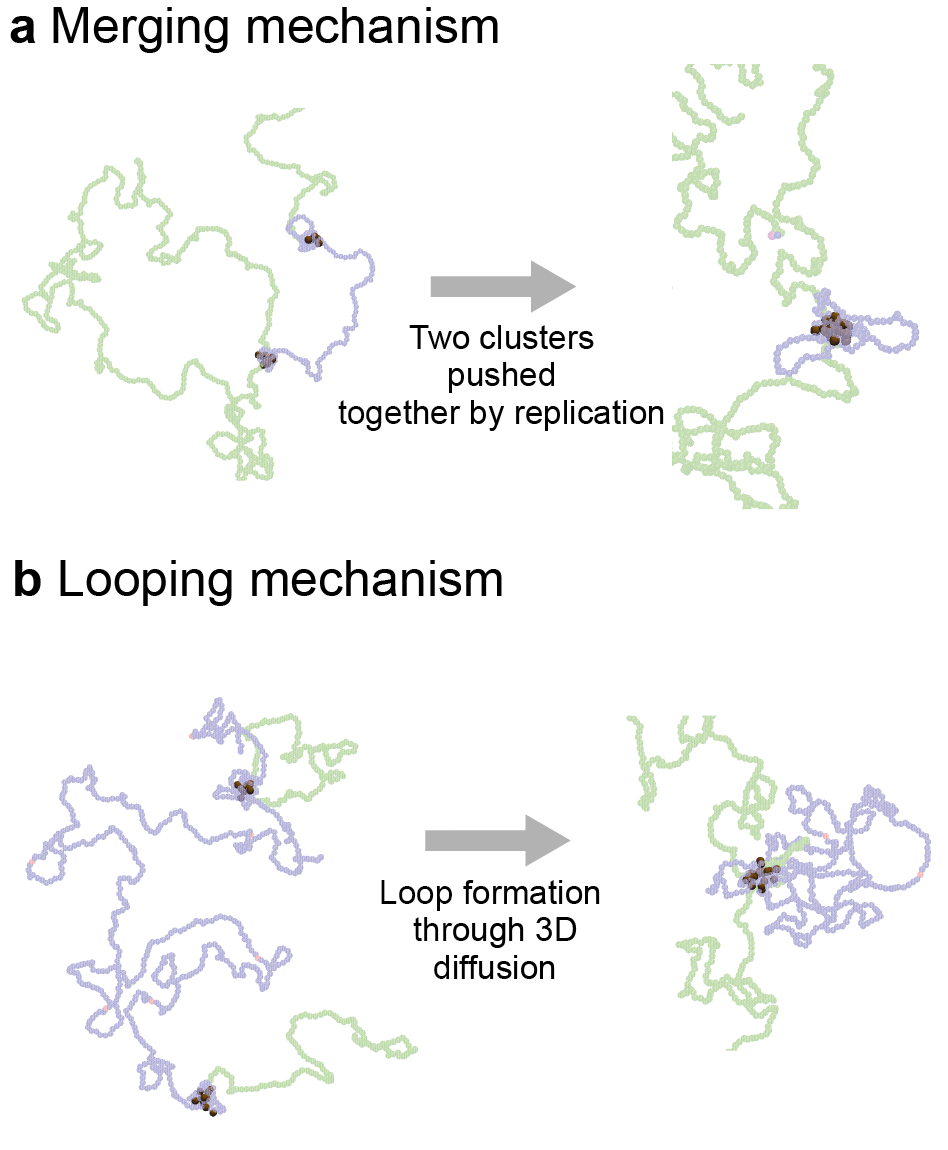} \caption{\textbf{Snapshots illustrating two mechanisms leading to growth of FF clusters.} \textbf{a} Merging mechanism: two FF clusters, pushed by the replication process, travel along the chromatin filament getting closer to each other (left panel). Eventually, they meet and merge, forming a bigger cluster and a few short-range chromatin loops (right panel, see the two small blue loops as an example of short-range chromatin loops). Beads composing the chromatin filament are represented as transparent to highlight FF clusters. \textbf{b} Looping mechanism: two FF clusters are far apart along the contour length of the chromatin filament (left panel), but Langevin dynamics bring them close in $3D$ resulting in their merging and in the formation of long-ranged loops (right panel). }\label{Fig:growth_mechanisms}
\end{figure}

Above we observed that fork and FF clusters tend to become bigger during replication. While this phenomenon is experimentally observed too, the mechanisms behind it are not completely clear yet~\cite{Chagin2016}. We now investigate how clusters grow, focusing, in particular on FF clusters.  In the $10$-origin system, two mechanisms for the growth of FF clusters can be identified, and are illustrated using snapshots from two different simulations. The first, a \textit{merging mechanism}, is shown in Fig.~\ref{Fig:growth_mechanisms}a. Here, two FF clusters track along the template chromatin filament pushed by replication itself (Fig.~\ref{Fig:growth_mechanisms}a,left panel) and, when they are close along the filament, they merge forming short-range chromatin loops (Fig.~\ref{Fig:growth_mechanisms}a, right panel). The second, an unexpected \textit{looping mechanism} (Fig.~\ref{Fig:growth_mechanisms}b), involves two clusters, far along the chromatin strand, which get closer in $3D$ through diffusion and merge (due to the increase of chromatin binding sites) to form a single bigger cluster. This is also associated with the formation of long-ranged chromatin loops.\\
Combined, these mechanisms mimic the enlargement of foci and their movement along the genome seen in vivo~\cite{Nakamura1986, Vouzas2021}. While the merging mechanism has been observed and predicted previously~\cite{Meister2007}, effects on looping have yet to be tested.
Notably, our simulations§ also suggest that FF clusters can increase in size by following an \textit{unbinding-rebinding} mechanism where some FFs, belonging to a cluster, abandon the chromatin filament to bind it again in the proximity of another cluster (see Fig.~S2). Note, however, that this phenomenon does not correspond to the growth of replication factories observed experimentally, as it does not increase the underneath cluster of forks.

\subsection*{The role of Rif1 in fission yeast replication}

\begin{figure*}
    \centering
    \includegraphics[width=1\textwidth]{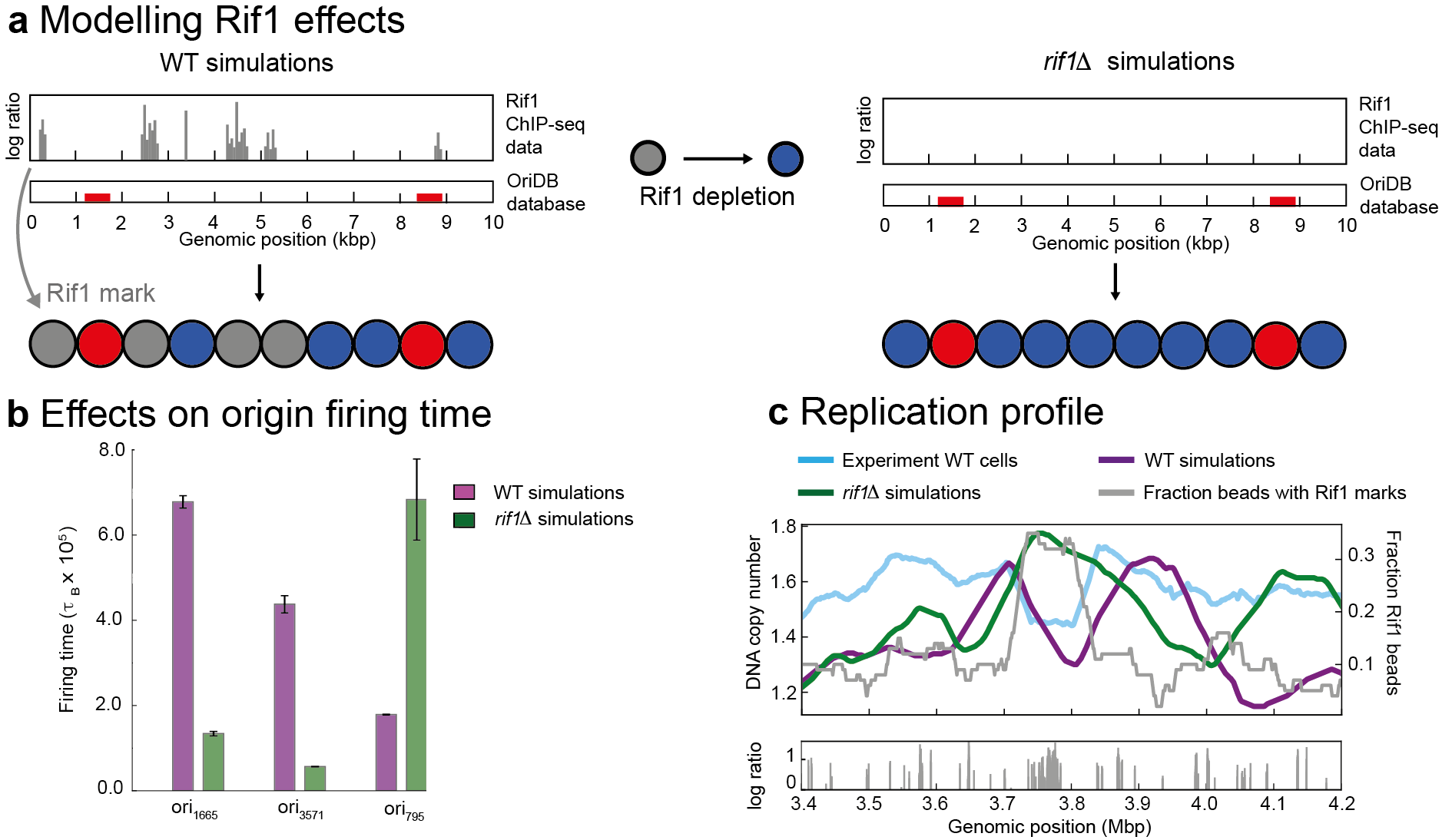} \caption{\textbf{Simulating the role of Rif1 in replication in fission yeast.}  \textbf{a} Cartoon of models for simulations including Rif1 (WT simulations, on the left), and simulations for Rif1 depleted cells (\textit{rif1}$\Delta$ simulations, on the right). The top panels, showing Rif1 ChIP-seq data and origin positions, do not correspond to real data but are only used to explain how bead types are introduced in the model. Beads are $1 \, kbp$ large and they have the Rif1 mark if they cover a region where there is at least one Rif1 ChIP-seq peak (grey beads in the left panel). Beads with Rif1 marks have a purely steric interaction with FFs. Origins are obtained from the OriDB database~\cite{OriDB}; when a bead contains both Rif1 peaks and an origin (as the second to last bead in the left panel), that bead is considered as a replication origin.  
    Following Rif1 depletion, beads with Rif1 marks are converted to conventional unreplicated chromatin beads which are weakly sticky FFs (blue beads, see right panel). \textbf{b} The firing times for three different origins of chromosome $1$  is plotted for $WT$ and \textit{rif1}$\Delta$ simulations. $ori_{1665}$ is a late origin in $WT$ simulations, which becomes a early origin in \textit{rif1}$\Delta$ conditions (the change is statistically significant, the corresponding $p$-value is $<10^{-12}$). The firing time of the early origin $ori_{3571}$ decreases too following depletion ($p<10^{-12}$). Instead, the origin $ori_{795}$, which fires early in $WT$ conditions, becomes a late origin in \textit{rif1}$\Delta$ conditions ($p\sim 1.1 \times 10^{-7}$). \textbf{c} Top panel: replication curves for the centromeric and pericentromeric region of chromosome $1$. $WT$ simulations produce a replication profile (purple curve) which significantly correlates with the experimental profile (light blue curve). Both these profiles anti-correlate with the fraction of beads with Rif1 marks (grey curve). Simulations of  \textit{rif1}$\Delta$ (green curve), instead, do not show any correlation with experiments. The three replication profiles (light blue, purple, and green curves) refer to the left y-axis, while the fraction of beads with Rif1 marks refer to the right y-axis. All four curves are obtained with a running average,  using a window whose size is $100 \, kbp$. Bottom panel: experimental Rif1 ChIP-seq data for the same region of chromosome $1$ analysed in the top panel.}\label{Fig:Rif1}
\end{figure*}

Rif1 was originally discovered as a telomere-binding protein in budding yeast~\cite{Hardy1992}; later, it was found to be highly conserved and to play a critical role in the first steps of origin firing in various eukaryotes~\cite{Hayano2012, Mattarocci2016}. For example, in fission yeast cells bearing a Rif1 deletion (\textit{rif1}$\Delta$), certain origins fire either earlier or later compared to those in the wild-type ($WT$), indicating that Rif1 affects origin firing probabilities. However, the mechanisms behind the down-regulation or up-regulation of firing times is not known yet.

In this section we introduce Rif1 in our model with the aim of explaining  different origin firing times in WT and \textit{rif1}$\Delta$ fission yeast cells as a consequence of a change in the interactions between chromatin and FFs.  

To set up our simulations, the following procedure is used (see Materials and Methods for more details). Rif1 binding sites are first identified, through the ChIP-seq peaks published in Ref.~\cite{Hayano2012}. Then,  each chromosome is coarse grained into $1$ kbp beads. In each $kbp$ there might be multiple Rif1 ChIP-seq peaks. A chromatin bead is endowed with the Rif1 mark if, in the $1$ kbp genetic region it covers, there is at least one Rif1 ChIP-seq peak (Fig.~\ref{Fig:Rif1}a, left panel). As Rif1 is often located close to heterochromatin and late or dormant origins~\cite{Yamazaki2013}, we assume that beads containing the Rif1 mark (grey beads in Fig.~\ref{Fig:Rif1}a) are non-binding for FFs. 
Finally, the position of fission yeast replication origins is obtained from the \textit{OriDB} database~\cite{OriDB} (see Materials and Methods for details and Table S1 for the complete list of origins). If a chromatin bead contains both Rif1 ChIP-seq peaks and a replication origin,  that bead is assumed to simply be an origin without introducing any Rif1 mark (see Fig.~\ref{Fig:Rif1}a, left panel).  
As in experiments, we aim to investigate the change in the replication timing following Rif1 depletion (\textit{rif1}$\Delta$). The latter is modelled by transforming beads with Rif1 marks into conventional unreplicated chromatin beads, which stick weakly to FFs (see Fig.~\ref{Fig:Rif1}a, right panel, and Fig.S3). $10$ independent simulations are performed in WT  (i.e. introducing Rif1 marks) and in \textit{rif1}$\Delta$ conditions. The following results refer to chromosome $1$, the longest of the $3$ chromosomes composing the fission yeast genome.

The first step consists of comparing the effects of \textit{rif1}$\Delta$ in simulations with those found experimentally. For simplicity, we define early origins as those origins which, on average, fire in the first half of the replication process, while late origins fire in the second half. [As the average total replication time is equal to $T_{rep} \sim 1.3 \times 10^6 \, \tau_B$, early origins fire at times $0 \lesssim t \lesssim 6.5  \times 10^5 \, \tau_B$ and late origins at times $6.5 \times 10^5 \, \tau_B \lesssim t \lesssim 1.3 \times 10^6 \, \tau_B$.] In Fig.~\ref{Fig:Rif1}b, we plot the average firing time for three origins, $ori_{1665}$, $ori_{3571}$ and $ori_{795}$. In $WT$ simulations, $ori_{1665}$ is a late origin (its firing time is $\sim 6.8 \times 10^5 \, \tau_B$), while $ori_{3751}$ and $ori_{795}$ are early origins. It can be observed that, as in experiments, Rif1 depletion affects the firing time both positively and negatively: while the firing time for $ori_{1665}$ and $ori_{3751}$ decreases in \textit{rif1}$\Delta$ simulations, the firing time of $ori_{795}$ increases and the latter becomes a late origin in Rif1 depleted conditions.

We suggest that these contrasting effects can be explained by considering how interactions change upon Rif1 depletion. When Rif1 is knocked-out, beads with Rif1 marks become weakly (non-specifically) sticky for FFs. This provokes two effects. First, wild-type late origins surrounded by $Rif1$ beads are more likely to fire earlier in \textit{rif1}$\Delta$ simulations as FFs can bind nearby sites and hence find the origins more easily. Second, the conversion of Rif1-marked beads into weakly sticky sites changes the distribution of FF binding sites along the chromatin filament and its 3D structure. Wild-type origins (both early and late), far from extended $Rif1$ domains can fire earlier when $Rif1$ is depleted, but they can also fire later, depending on how the near binding sites disposition and the chromatin 3D structure have been altered.
Interestingly, in \textit{rif1}$\Delta$ simulations,  the total replication time is reduced to $T_{rep} \sim 9 \times 10^5 \, \tau_B$, in line with what experimentally observed~\cite{Hayano2012}. 

The effects of $Rif1$ depletion are also visible in the replication profile -- the graph showing the average quantity of DNA for each chromatin site during the replication process. Fig.~\ref{Fig:Rif1}c portrays replication profiles for the centromere and pericentromeric region in chromosome $1$, where the bound Rif1 is abundant. The grey curve (referred to y-axis on the right) shows, for each bead, the fraction of beads with Rif1 marks. As the presence of Rif1 usually delays firing, the experimental replication profile for WT cells (light blue curve, referring to the left y-axis) assumes its minimum values where the fraction of beads with Rif1 marks is maximum. Our WT simulations (purple curve, left y-axis) show the same behaviour (the correlation between WT simulations and experiments in WT cells is $0.34$ with $p-value<10^{-10}$).
As expected, the replication profile for \textit{rif1}$\Delta$ simulations (green curve, left y-axis) does not correlate with experiments. 

In conclusion, Rif1 effects on replication of fission yeast cells can be captured by our $3D$ model which highlights how non-specific interactions can substantially change the probability of firing, as they facilitate the search for origins when they are present, and affect the binding landscape of FF on chromatin, as well as chromatin context.

\section*{Discussion}

In summary, here we have proposed a 3D model for chromatin replication and characterised its emergent behaviour. This model is fundamentally different with respect to most previous DNA replication models which are effectively 1-dimensional~\cite{Lob2016, De2010, Kelly2019}. Unlike previous 3D models for DNA replication~\cite{Jun2006}, ours focuses on chromatin rather than bacteria and explicitly includes active firing factors -- which model generic complexes of replisome components, such as DNA polymerases and helicases. This allows us to study the effect of different types of chromatin-protein interactions, such as the balance between non-specific attraction between firing factors and unreplicated chromatin, and the specific attraction to origins and forks. We also model replication dynamics, so that we can ask questions on the time evolution of 3D chromatin and protein structures at mobile forks. As we discuss below, besides recapitulating known features of chromatin replication, our model allows us to make definite mechanistic predictions, which could be experimentally testable. 

Our main result is that clusters of firing factors and forks spontaneously form during replication: these clusters diffuse slowly while extruding loops of replicated chromatin. The extrusion of replicating chromatin loops is qualitatively consistent with the biological models of immobile replisomes~\cite{Yuan2019} and of replication factories~\cite{Cook1999}; another related biophysical model is that of loop extrusion via SMC proteins, although in our case clustering of firing factors and forks is required to extrude replication loops, so that extrusion is an emergent property of the model. More specifically, we predict that extrusion requires two main ingredients: (i) a motor activity of firing factors at replication forks, which is natural to assume as they model complexes of molecular motors such as DNA polymerases and helicases, and (ii) cluster formation. The latter occurs through an active generalisation of bridging-induced phase separation (BIPS)~\cite{Brackley2013, Ryu2021}, which stands for the generic tendency of multivalent proteins interacting with chromatin to cluster. 

BIPS in our context essentially requires non-specific interactions to occur, and indeed when abrogating them extrusion is not observed in the simulations, and instead replisomes separately track on chromatin. Non-specific interactions are likely important {\it in vivo}, and BIPS may underlie the formation of clusters of pre-replication complexes~\cite{Li2023}.
Before replication, BIPS creates microphase-separated aggregates~\cite{Brackley2016} due to the combination of non-specific attraction to non-replicated chromatin and specific interactions to the origins. These aggregates later nucleate sites where replication initiates mimicking what happens {\it in vivo}, where transcriptional hubs (or factories) co-localise with sites of replications (or early replication factories)~\cite{Hassan1994}. 

Our dynamic model can be used to study the morphology and dynamics of clusters of firing factors and forks. Regarding morphology, we observe that clusters typically involve a significantly larger number of firing factors with respect to forks. Concerning the dynamics, we observe a non-monotonic behaviour, where clusters first grow, then shrink as replication terminates. This is qualitatively similar to what was observed in cells~\cite{Hozak1994}.

Additionally, inspection of the dynamical trajectories of our model allows us to identify all the kinetic events through which replication clusters may grow or evolve in S-phase. First, forks or replisomes may collide and merge. 
Second, the trajectories show a distinct mechanism through which replication factories that are far apart along the chromatin colocalise in space: this is via the formation of a long-range chromatin loop. We speculate that this fully $3D$ mechanism would ignite the firing of an origin by forming a chromatin loop between the inactive origin itself and a replication cluster.  It would be interesting to seek evidence of this looping-mediated origin activation in the future, possibly by analysing correlations between data on origin activities over time and Hi-C maps of chromosome contacts in the S-phase.

Finally, we quantitatively compare the replication patterns predicted by our 3D model with those found experimentally in fission yeast \textit{S. Pombe}, where replication timing is understood to be in large part determined by the Rif1 protein, which inactivates chromatin and slows down replication. Our results show that chromatin context and local 3D structure are important for the timing or origin activation, and we suggest that Rif1 activity can be modelled by abrogating the non-specific binding of unreplicated chromatin to firing factors. While in a 1D model this non-specific interaction would be inconsequential, in our model the local absence of non-specific binding tends to inhibit origin activation, in line with experimental results. Strikingly, our simulations also predict that knocking out Rif1 does not always upregulate origin activity through the elimination of heterochromatin, but can substantially dial down the activity of some previously early-replication origins, due to the appearance of new competitors (which were previously inhibited by Rif1). This effect matches experimental observations of subtle and non-trivial changes in the replication timing of origin after Rif1 knockout.

Besides providing interesting results about the formation and dynamics of clusters formed by firing factors and forks, our model also gives interesting insights on potential mechanisms to avoid the formation of stalled forks.  
Even if our model does not include transcription-related molecules or DNA breaks, it still predicts the formation of temporarily inactive forks where thermal noise leads to factor disengagement from a fork. 
The corresponding continuous binding and unbinding of FFs predicted by our model is in line with the process of replication of Okazaki fragments~\cite{Ogawa1980}, and also with recent experiments where components of yeast replisomes are observed only to be transiently bound to replication forks~\cite{Beattie2017}. 
In our simulations, these temporarily inactive structures can be readily rescued,
as the weak attraction between unreplicated chromatin and FFs facilitates the reassembly of an FF cluster close to them. This may avoid the formation of permanently stalled forks, which would instead biologically require the DNA damage response to be reactivated.  

More generally, we note that non-specific attraction might act not only between forks and FFs, but also between forks and biomolecules that are known to be involved in repairing stalled forks such as the enzyme RecG which is needed in order to restart replication of temporarily inactive forks in the \textit{Escherichia Coli} genome~\cite{Bianco2021}. 
In the future, it would be interesting to experimentally investigate whether the reactivation of such forks is easier when these are embedded in unreplicated euchromatin, due to non-specific interactions, as predicted by our model. \\ 


\if{
\begin{figure}[htbp]
    \centering
    \includegraphics[width=1\columnwidth]{Figures/Figure7.png}
    \caption{\textbf{Formation of stalled replication forks in simulations.} In the multi-origin set-up, a FF cluster can form in proximity of a replication origin which fires forming two replication forks. While at the beginning the two forks replicate together resulting in the extrusion of replicated chromatin (panel on the left), the presence of thermal noise (due to other origins and forks) can lead to the displacement of the FF cluster. A possible resulting scenario is one in which the cluster remains close to one of the two original forks, while the other fork stops replicating as no FFs are close by (panel on the right). This process converts an extrudin cluster into a stalled fork and a tracking replisome. 
    }
    \label{Fig:stalled_forks} 
     
\end{figure}
}\fi


\subsection*{MATERIALS AND METHODS}

\subsection{Molecular dynamics simulations}

Chromatin is modelled as a semi-flexible chain of $N$ beads, each with diameter $\sigma$ = 15 nm (corresponding to $1 \, kbp$). Bonds between consecutive beads are treated as  harmonic springs 
\begin{equation}
V_{H}(r)= K_{H}(r-R_{H})^2\label{Eq:harmonic_pot},
\end{equation}
with typical spring length $R_{H}=1.1 \sigma$ and spring constant $K_{HA}=200 \, k_BT/\sigma^2$, where $k_B$ is the Boltzmann constant, and $T=300 \, K$ the temperature of the system.
The chain's stiffness is modeled by a Kratky-Porod potential:
\begin{equation}
    V_{B}(\phi) = K_{B}(1+\cos\phi)\label{Eq:kratky-porod},
\end{equation}
with $\phi$ being the angle between three consecutive beads, and $K_{B}$ the rigidity coefficient. The latter is set equal to $K_{B}=4 \, k_BT$ to give a persistence length  $l_p \sim 60 \, nm$ (compatible to that of chromatin~\cite{Langowski2006}).
The excluded-volume interaction between non-consecutive beads at spatial distance $r$ is ruled by the  Weeks-Chandler-Anderson (WCA) potential
\begin{equation}
    V_{WCA}(r) = 4k_BT \left[ \left( \frac{\sigma}{r} \right)^{12} - \left(\frac{\sigma}{r} \right)^6 +\frac{1}{4} \right]\Theta(2^{1/6} \sigma -r)\label{Eq:WCA_potential},
\end{equation}
The chain is in diluted conditions, immersed in a cubic simulation box of size $110 \, \sigma$ with freely diffusing brown FFs.  The initial configuration involves unreplicated chromatin sites and origins (blue and red beads in Fig.~\ref{Fig:model}a). After a pre-equilibration for a time $T_{eq,pol} = 1.5 \times 10^6 \, \tau_{LJ}$ (where $\tau_{LJ}$ is the Lennard-Jones time unit for simulations), FFs (brown spheres in Fig.~\ref{Fig:model}) are initially inserted in random positions into the volume. Then, a soft potential $V_{SOFT}$ is applied between them and beads in the chain for a time $T_{SOFT} = 10^3 \, \tau_{LJ}$ in order to displace those FFs that overlap beads in the chain. The soft potential is described by
\begin{equation}
    V_{SOFT}(r)=A\left[  1+ cos\left( \frac{\pi r}{r_c} \right)\right]\Theta(r-r_c)\label{Eq:soft_potential},
\end{equation}
where $A=100 k_BT$ describes the strength of the potential and $r_c=2^{1/6} \, \sigma$ is the threshold below which the potential is effective. Consequently, the system is further equilibrated by inserting only steric repulsions between  FFs and beads in the chain for an additional time $T_{steric}=10^3 \, \tau_{LJ}$.
Replication initiates (Fig.~\ref{Fig:model}B) after a time $T_{eq,tot}=T_{eq,pol}+T_{SOFT}+T_{steric}$ by switching on an attractive interaction between FFs and the chain that is described by a truncated and shifted Lennard-Jones potential:
\begin{equation}
V_{LJ/cut}(r) = \left[ V_{LJ}(r) - V_{LJ}(r_c) \right]\Theta(r_c-r),
\end{equation}
with 
\begin{equation}
V_{LJ}(r)=4\varepsilon \left[ \left( \frac{\sigma}{r} \right)^{12}- \left( \frac{\sigma}{r} \right)^6 \right].
\end{equation}
We consider a cutoff distance  $r_c=1.8 \, \sigma$, while the attraction strength is $\varepsilon_{origin}= 6 \, k_BT$ between  FFs and origins, and $\varepsilon_{ns} = 4 \, k_BT$ between FFs and unreplicated chromatin beads (except in specified cases where the last two parameters are changed to investigate extrusion of replicated chromatin, see Fig.~\ref{Fig:phase_diagram}).  If an origin at site $i$ has at least one  FF at a distance $r<r_c=1.8 \, \sigma$, it fires with probability $P_{fire}$ to create a pair of forks (black beads in Fig.~\ref{Fig:model}) that experience an attraction $\varepsilon_{fork}=10 \, k_BT$ with FFs.  In case both sites $i-1$ and $i+1$ are unreplicated chromatin sites or replication origins, the pair of forks is created in  $(i-1,i)$ or $(i,i+1)$ with equal probability. If $i+1$ (or $i-1$) is occupied by another pre-existing fork, the pair of forks is created in $(i-1,i)$ (or in ($i,i+1$)). Similarly, if $i$ corresponds to the first polymer bead, $i=1$, the forks are created in $(1,2)$, while if it corresponds to the last polymer bead $i=N$, the forks will be placed in $(N-1,N)$. Finally, if neither sites $i+1$ and $i-1$ are available, the pair of forks is not created. 
Once created, the two forks move step-wise along the chain independently and in opposite directions whenever a FF is located at a distance $d\leq 1.8 \, \sigma$ from them (Fig.~\ref{Fig:model}C). Supposing that a fork is in position $i$ and that a FF is close by, a replication step involves the fork moving, for instance, to the site $i+1$, while the site $i$ becomes replicated chromatin and a newly synthesised chromatin bead is inserted in the system (green beads in Fig.~\ref{Fig:model}). Both replicated segments are connected to the forks they are originated from through harmonic potentials as in Eq.~\ref{Eq:harmonic_pot} and they both also provide the same energy contribution: a Kratky-Porod potential (Eq.~\ref{Eq:kratky-porod}) to represent the chain's stiffness, a harmonic interaction (Eq.~\ref{Eq:harmonic_pot}) for bonded beads belonging to one of the two chains and a repulsive soft potential (Eq.~\ref{Eq:soft_potential}) to describe the interaction between green synthesised beads and any other non-bonded bead (FFs, chromatin beads from the same replicated segment and chromatin beads from other replicated or unreplicated segments). When two forks collide, they disappear to leave two conjoined strings of green replicated beads that cannot be re-replicated as they contain no replication origin and have no affinity for brown beads (Fig.~\ref{Fig:model}D). Once a fork reaches an extremity of the template chain and replicates it, the two newly replicated filaments are not longer join together and diffuse independently. Replication stops once every bead of the original template chain has been replicated.

The dynamics of a bead at position $\bf{r}_i$  is described by the Langevin equation:
\begin{equation}
m_i\frac{\partial^2 \bf{r}_i}{\partial t^2} = \bf{\nabla}_i U - \gamma_i \frac{\partial \bf{r}_i}{\partial t} + \sqrt{2k_BT\gamma_i}\bf{\eta}_i,
\end{equation}
where U is the total energy of the system and $\gamma_i$ the friction on the $i-th$ bead due to the solvent. The term $\bf{\eta}_i$ represents thermal noise whose components are such that:
\[
\langle \eta_{i\alpha}(t) \rangle=0 \; \; \text{and} \; \; \langle \eta_{i\alpha}(t) \, \eta_{j\beta}(t') \rangle=\delta_{ij}\delta_{\alpha\beta}\delta(t-t')
\]
where $\delta_{ij}$ and $\delta_{\alpha \beta}$ are the Kronecker delta and $\delta(t-t')$ is the Dirac delta. Simulations are performed using the software LAMMPS for Molecular Dynamics~\cite{Plimpton1995} (using a time step $dt= 0.01 \tau_{LJ}$) and an external C++ code called by the LAMMPS script. The C++ code is needed to implement fork movements and synthesis of the newly- replicated chain. The simulation time between two consecutive calls of code by a LAMMPS script, $T_{call}$, is then directly related to the fork velocity through the relationship $v_{fork} = \frac{1}{T_{call}} \, \sigma/\tau_{LJ}$ as a fork moves $1 \, \sigma$ between two consecutive calls of the code (in the presence of at least one nearby FF). To map fork velocity in real units, we first map the simulation time $\tau_{LJ} = \sigma \sqrt{m/k_BT}$. By considering the Brownian time $\tau_B = \sigma^2 / D_B$ (where $D_B$ is the diffusion coefficient) and the decorrelation time $\tau_{dec}=m/\gamma$ (where $\gamma$ is the friction of the solvent), we get $\tau_{LJ} = \tau_{dec} = \tau_{B}$ (as in our simulations we set $\sigma = \gamma = k_BT = m = 1$ and $D_B = k_BT/\gamma$). Therefore, it is possible to use $\tau_B$ to map time from simulation to real units. By employing the Stokes-Einstein equation $\gamma= 3 \pi \sigma \eta_{sol}$, we obtain $\tau_{LJ}=\tau_{B} = \frac{3\pi \sigma^3 \eta_{sol}}{k_BT}$ which, for $\sigma=1 \, kbp \sim 15 \, nm$, $T=300 \, K$ and $\eta_{sol} \sim 150 \, cP$, provides $\tau_{LJ} \sim 1 \, ms$. In our simulations we use $T_{call} = 3 \cdot 10^3 \, \tau_{LJ} - 5 \cdot 10^3 \, \tau_{LJ}$ resulting in $v_{fork} \sim 12 - 20 \, kbp/min$ that is relatively close to the average fork velocity in eukaryotes~\cite{Gispan2017}.

Different initial set-ups are used. In the 1-origin case, the chain contains $1000$ beads (representing $1 \, Mbp$), with a single replication origin placed in the middle (bead index $501$; beads numbered from $1$ to $1000$). In the 10-origin case, the chain contains $1000$ beads with $10$ equally-spaced origins at beads $50, \, 150, \, 250....950$ (so origins are $100 \, kbp$ apart). When considering chromosome 1 in fission yeast (\textit{S. Pombe}), a chain of 5,580 beads represents $5.58 \, Mbp$~\cite{Wood2002}, with origins placed accordingly to OriDB~\cite{OriDB} database and including only \textit{likely} and \textit{confirmed} origins (Table Si gives the complete list).

\subsection{Rif1 ChIP-seq data analysis}
Positions of Rif1 binding sites were extracted from published ChIP-seq data~\cite{Hayano2012}, where the authors accepted a peak if three conditions applied: (i) the \textit{test p-value} has to be $<0.025$, (ii) the \textit{p-value change} has to be $<0.001$, and (iii) the peak must be in a cluster of peaks satisfying the two previous conditions. We enforce the last condition by requiring the two nearest-neighbour peaks to respect the requirements about the \textit{test p-value} and the \textit{p-value change}. After selecting ChiP-seq peaks, a Rif1 mark is applied to each bead if the coarse-grained 1 kbp region contains at least one ChIP-seq peak.

\section{Acknowledgements}
We thank 
C.~A. Brackley and D. Jost for useful discussions. This work was supported by the Wellcome Trust (223097/Z/21/Z). E.O. acknowledges support from grant PRIN 2022R8YXMR funded by the Italian Ministry of University and Research.

\section{Author contributions}
G.F. performed simulations. G.F., D.M. and E.O. analysed the data and designed the research. All authors contributed to writing the manuscript.

\section{Competing interests}
The authors declare no competing interests.

\bibliography{sn-bibliography}


\end{document}